\documentclass[conference]{IEEEtran}
\IEEEoverridecommandlockouts
\usepackage{cite}
\usepackage{amsmath,amssymb,amsfonts}
\usepackage{algorithmic}
\usepackage{graphicx}
\usepackage{textcomp}
\usepackage{xcolor}
\def\BibTeX{{\rm B\kern-.05em{\sc i\kern-.025em b}\kern-.08em
    T\kern-.1667em\lower.7ex\hbox{E}\kern-.125emX}}
\begin{document}

\title{Enhancing ASR Performance in the Medical Domain for Dravidian Languages \\
% {\footnotesize \textsuperscript{*}Note: Sub-titles are not captured for https://ieeexplore.ieee.org  and
}

\author{
\IEEEauthorblockN{
Sri Charan Devarakonda\IEEEauthorrefmark{1},
Ravi Sastry Kolluru\IEEEauthorrefmark{1},
Manjula Sri Rayudu\IEEEauthorrefmark{2},\\
Rashmi Kapoor\IEEEauthorrefmark{2},
Madhu G\IEEEauthorrefmark{2},
Anil Kumar Vuppala\IEEEauthorrefmark{1}
}
\IEEEauthorblockA{\IEEEauthorrefmark{1}
Language Technologies Research Centre (LTRC)\\
International Institute of Information Technology Hyderabad (IIIT-H) ,
Hyderabad, India \\
\{sricharan.d, kolluru.s\}@research.iiit.ac.in,
anil.vuppala@iiit.ac.in
}
\IEEEauthorblockA{\IEEEauthorrefmark{2}
Vallurupalli Nageswara Rao Vignana Jyothi Institute of Engineering and Technology (VNRVJIET) , Hyderabad, India\\
\{manjulasree\_r, rashmi\_k, madhu\_g\}@vnrvjiet.in
}
}

\maketitle

\begin{abstract}
Automatic Speech Recognition (ASR) for low-resource Dravidian languages like Telugu and Kannada faces significant challenges in specialized medical domains due to limited annotated data and morphological complexity. This work proposes a novel confidence-aware training framework that integrates real and synthetic speech data through a hybrid confidence mechanism combining static perceptual and acoustic similarity metrics with dynamic model entropy. Unlike direct fine-tuning approaches, the proposed methodology employs both fixed-weight and learnable-weight confidence aggregation strategies to guide sample weighting during training, enabling effective utilization of heterogeneous data sources. The framework is evaluated on Telugu and Kannada medical datasets containing both real recordings and TTS-generated synthetic speech. A 5-gram KenLM language model is applied for post-decoding correction. Results show that the hybrid confidence-aware approach with learnable weights substantially reduces recognition errors: Telugu Word Error Rate (WER) decreases from 24.3\% to 15.8\% (8.5\% absolute improvement), while Kannada WER drops from 31.7\% to 25.4\% (6.3\% absolute improvement), both significantly outperforming standard fine-tuning baselines. These findings confirm that combining adaptive confidence-aware training with statistical language modeling delivers superior performance for domain-specific ASR in morphologically complex Dravidian languages.

\end{abstract}


\section{Introduction}

Recent advances in ASR for low-resource languages have leveraged deep learning and synthetic data augmentation to address data scarcity. Synthetic data generation has shown promising results, with~\cite{b1} achieving 5–10\% relative WER improvements using integrated text-to-mel-spectrogram generators. The IndicWav2Vec project~\cite{b2} developed multilingual self-supervised models for Indian languages including Telugu, while others explored cross-lingual and multilingual pretraining. More recently,~\cite{b3} proposed a domain adaptation framework for ASR using only synthetic data, demonstrating the potential of synthetic approaches in addressing data limitations. Similarly, generative error correction has been explored in the post-processing stage, with~\cite{b4} showing that synthetic data and retrieval-augmented correction can further enhance robustness, particularly in out-of-domain scenarios. Complementary work such as~\cite{b5} showed that careful design of TTS- and VC-based augmentation can reduce WER by up to 35\% relative, emphasizing that naive mixing of synthetic and real data is suboptimal.

Automatic Speech Recognition (ASR) systems for low-resource languages face significant challenges due to limited training data, particularly in specialized domains such as healthcare. The scarcity of annotated medical speech data in Telugu further exacerbates these issues, resulting in poor performance when general-purpose ASR models are applied to medical contexts. Telugu, being a morphologically rich language with complex phonetic structures, presents additional challenges for ASR development. These challenges are addressed through a comprehensive framework that combines synthetic data generation, confidence-aware training, and sophisticated post-processing techniques to develop robust ASR systems for Telugu medical speech recognition.

Medical domain ASR presents further challenges due to specialized terminology and pronunciation variability. For Indian languages, text-to-speech synthesis has been advanced by unified frameworks for dataset collection~\cite{b6} and generative flow-based models like Glow-TTS~\cite{b7}, though their integration into ASR training pipelines remains limited. While robust multilingual ASR systems like Whisper~\cite{b8} show strong general performance, they primarily focus on English, leaving Indian language support limited, particularly in specialized domains like healthcare. \cite{b9} introduced Modality Confidence Aware Training (MCAT), dynamically weighting audio and text modalities based on estimated confidence scores. To address noisy or uncertain data, confidence-aware training strategies have gained traction in recent years. Other approaches have combined multi-stage training with synthetic speech and incorporated strategies like parameter regularization and weighted sampling.

Entropy-based methods~\cite{b10} aggregate per-frame uncertainty for CTC and RNN-T~\cite{b11} models, significantly improving error detection. Recent research has also explored advanced confidence estimation methods that go beyond simplistic maximum-probability scores. TeLeS~\cite{b12} introduces a Temporal-Lexeme Similarity score that captures alignment and partial errors, while TruCLeS~\cite{b13} combines true class probability with lexical similarity for finer-grained confidence supervision. In parallel, Whisper-based confidence estimation~\cite{b14} fine-tunes large ASR models to directly output word-level confidence, outperforming conventional lightweight Confidence Estimation Modules (CEMs) across domains. Together, these works indicate a shift toward adaptive, linguistically informed, and model-integrated confidence scoring.

However, most existing methods rely on static mixing ratios or simplistic confidence metrics rather than adaptive, multi-dimensional quality assessments. The main contributions of this work are: 
\begin{itemize}
    \item A curriculum learning strategy that evolves confidence weighting during training,
    \item Proposed a novel hybrid confidence scoring mechanism that combines static perceptual metrics with dynamic model entropy for data quality assessment,
    \item A two-stage post-processing pipeline utilizing both statistical and neural language models for transcript correction. The proposed approach leverages both statistical and neural language models to improve transcript quality, providing a comprehensive solution for Telugu medical ASR that addresses the unique challenges of morphologically rich languages in specialized domains. 
\end{itemize}

\section{Proposed Methodology}

\begin{figure}[h]
\centering
\includegraphics[width=0.45\textwidth]{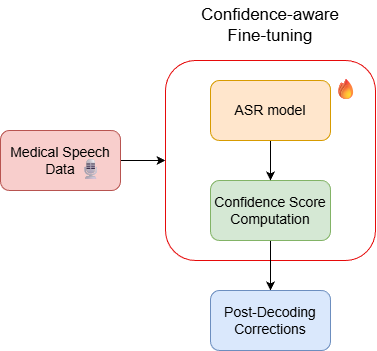}
\caption{Confidence-aware training pipeline for Telugu medical ASR.}
\label{fig:data_flow}
\end{figure}

This section explains how the proposed confidence-aware training framework enhances Telugu medical speech recognition systems. The training pipeline, illustrated in Figure~\ref{fig:data_flow}, evaluates the reliability of synthetic data, combines multiple confidence measures, fine-tunes the model on the most trustworthy examples, and applies post-decoding corrections to improve accuracy.

\subsection{ASR model}
We use both the Wav2Vec2~\cite{b15} and Whisper~\cite{b16} models for domain adaptation in Telugu medical ASR to analyze whether differences in pretraining strategies affect performance during fine-tuning. In particular, we study how the choice of architecture and training objective like CTC-based decoding~\cite{b17} in Wav2Vec2 versus sequence-to-sequence decoding in Whisper impacts adaptation effectiveness.

\subsection{Confidence Score Computation}

We propose a three-component confidence scoring mechanism to evaluate speech data quality in a heterogeneous dataset containing both real and synthetic speech. These confidence measures help prioritize reliable samples during training, enabling more effective utilization of mixed data sources for ASR model adaptation.

\subsubsection{Static Confidence Metrics}

\begin{itemize}

\item \textbf{Perceptual Score (All Data):}  
Acoustic features are extracted from both real and synthetic speech to assess speech quality and consistency, following approaches commonly used in synthetic speech evaluation~\cite{b18}. The extracted features include spectral centroid ($F_{sc}$), spectral rolloff ($F_{sr}$), mean MFCC ($F_{mfcc}$), pitch variation ($F_{pv}$), and energy ($F_e$). Each feature is normalized using min-max normalization:

\begin{equation}
\tilde{F}_i = \frac{F_i - F_i^{\min}}{F_i^{\max} - F_i^{\min}}, \quad i \in \{sc, sr, mfcc, pv, e\}
\label{eq:minmax_normalization}
\end{equation}

The perceptual score is computed as the mean of the normalized feature values:

\begin{equation}
S_{\text{perceptual}} =
\frac{1}{5}
(\tilde{F}_{sc} + \tilde{F}_{sr} + \tilde{F}_{mfcc} + \tilde{F}_{pv} + \tilde{F}_e)
\label{eq:perceptual_score}
\end{equation}

\item \textbf{Acoustic Similarity Score (Aligned Synthetic Data):}  
For synthetic speech aligned with real speech having identical transcripts, cosine similarity between MFCC feature representations is computed to measure acoustic similarity:

\begin{equation}
S_{\text{sim}} =
\cos(\text{MFCC}_{real}, \text{MFCC}_{synth})
\label{eq:similarity_score}
\end{equation}

\item \textbf{WER Score (Aligned Synthetic Data):}  
Synthetic audio is transcribed using an IndicWav2Vec model and compared with the ground-truth transcript shared with the real speech sample. The resulting confidence score is defined as:

\begin{equation}
S_{\text{wer}} =
1 - \text{WER}(\text{real}_{txt}, \text{synth}_{txt})
\label{eq:wer_score}
\end{equation}

\end{itemize}

\subsubsection{Per-Source Confidence Aggregation}

Final confidence scores are calculated for each data source based on the available metrics, as summarized in Table~\ref{tab:confidence_scores}. Two aggregation strategies are explored for combining the static confidence components.

\textbf{Fixed Weights}

\begin{equation}
C_{\text{static}} =
\alpha S_{\text{perceptual}} +
\beta S_{\text{sim}} +
\gamma S_{\text{wer}}
\label{eq:static_score}
\end{equation}

subject to

\[
\alpha + \beta + \gamma = 1, \quad
\alpha,\beta,\gamma \ge 0
\]

\textbf{Learnable Weights}

\begin{equation}
\begin{bmatrix}
\alpha \\
\beta \\
\gamma
\end{bmatrix}
=
\frac{1}{e^{w_1}+e^{w_2}+e^{w_3}}
\begin{bmatrix}
e^{w_1} \\
e^{w_2} \\
e^{w_3}
\end{bmatrix}
\label{eq:learnable_weights}
\end{equation}

where $w_1, w_2, w_3 \in \mathbb{R}$ correspond to the learnable parameters associated with $\alpha$, $\beta$, and $\gamma$.

\begin{equation}
C_{\text{learnable}} =
\alpha S_{\text{perceptual}} +
\beta S_{\text{sim}} +
\gamma S_{\text{wer}}
\label{eq:learnable_score}
\end{equation}

\begin{table}[t]
\centering
\caption{Confidence score availability for different data sources.}
\begin{tabular}{lccc}
\hline
\textbf{Source} & \textbf{Perceptual} & \textbf{Acoustic} & \textbf{WER} \\
 & \textbf{Score} & \textbf{Similarity} & \textbf{Score} \\
\hline
Real & Computed & 1.0 & 1.0 \\
Synthetic Aligned & Computed & Computed & Computed \\
Synthetic Unaligned & Computed & 0 & 0 \\
\hline
\end{tabular}
\label{tab:confidence_scores}
\end{table}

\subsubsection{Hybrid Confidence}

The proposed hybrid confidence mechanism combines static confidence scores with dynamic model-based confidence during training. Model uncertainty is estimated using entropy:

\begin{equation}
H = -\sum_{i} p(x_i) \log p(x_i)
\end{equation}

where $p(x_i)$ denotes the output probability distribution of the ASR model.

Model confidence is computed as:

\begin{equation}
C_{\text{model}} =
1 - \mathrm{normalize}(H)
\end{equation}

The final confidence score integrates both static and dynamic components:

\begin{equation}
C_{\text{final}} =
\lambda C_{\text{static or learnable}}
+
(1-\lambda) C_{\text{model}}
\end{equation}

where $\lambda$ is initialized at 1.0 to prioritize static confidence during early training and gradually annealed to 0.5 to balance both confidence sources.

\subsection{Confidence-Aware Fine-Tuning}

To incorporate confidence scores during training, the loss function is weighted by the final confidence value:

\begin{align}
\mathcal{L}_{\text{weighted}} &=
C_{\text{final}} \cdot \mathcal{L}_{\text{CE}}
\label{eq:weighted_loss} \\
\mathcal{L}_{\text{total}} &=
\frac{1}{N}\sum_{i=1}^{N} \mathcal{L}_{\text{weighted}_i}
\label{eq:total_loss}
\end{align}

where $N$ is the batch size and $\mathcal{L}_{\text{CE}}$ denotes the cross-entropy loss. This formulation ensures that high-confidence samples contribute more strongly to model updates.

\subsection{Post-Decoding Correction}

To further improve transcription accuracy and correct domain-specific recognition errors, both statistical and neural language models are applied during post-processing. A KenLM~\cite{b20} n-gram language model trained on domain-specific medical Telugu text captures local linguistic patterns and helps correct frequent ASR decoding errors. In addition, neural language models such as IndicBART~\cite{b21} and mT5~\cite{b22} refine the generated transcripts using contextual language understanding, improving overall transcription quality.

\section{Experimental Setup}

\subsection{Medical Domain Database}
\begin{figure}[h!]
\centering
\includegraphics[width=0.4\textwidth]{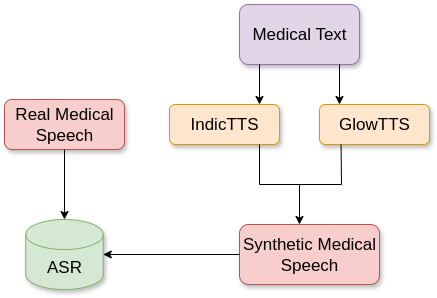}
\caption{Data sources and processing flow for the Telugu and Kannada medical ASR framework.}
\label{fig:pipeline_overview}
\end{figure}

Figure~\ref{fig:pipeline_overview} shows the database creation for this study. Given the difficulties associated with collecting annotated speech data, we used two different TTS models, IndicTTS~\cite{b23} and GlowTTS, to generate synthetic speech data in addition to in-house domain-specific speech data.

\subsubsection{Telugu In-House Data}
We used an in-house Telugu medical domain speech database consisting of 30 hours of read-mode recordings collected from 87 speakers, including 40 male and 47 female participants. The medical text was first normalized and then segmented into meaningful phrases that include medical terminology, disease names, and commonly used clinical expressions. All recorded audio samples were downsampled to 16 kHz to ensure consistency and compatibility with ASR models. Out of the 30 hours, 20 hours are used for training and the remaining 10 hours are reserved for testing.

\subsubsection{Telugu Synthetic Data}
The synthetic data consists of 40 hours generated through a multi-stage approach. 9.6 hours were produced using IndicTTS with text corresponding to the first 10 hours of real data, another 9.7 hours were generated using GlowTTS (trained separately on 11 hours of Telugu IndicTTS data) with text from the second 10 hours of real data, and the remaining 20.5 hours of synthetic speech were created from additional sentences not aligned with the real speech data, using both IndicTTS and GlowTTS systems, as shown in Figure~\ref{fig:pipeline_overview}. All data collection followed proper ethical guidelines with participant consent.

\subsubsection{Kannada Medical Data}
To further evaluate the framework, we extended the experiments to the Kannada medical domain. A total of 30 hours of Kannada data was prepared, consisting of 10 hours of real and 20 hours of synthetic speech data. The real speech data was obtained from publicly available Kannada medical datasets on Kaggle, which include domain-specific utterances and clinical terminology. The synthetic portion was generated using both IndicTTS and GlowTTS systems, with 10 hours each. The GlowTTS model was specifically trained using a Kannada dataset derived from the Indic corpus, similar to the approach used for the Telugu GlowTTS training. All Kannada datasets were normalized and downsampled to 16 kHz to maintain uniformity with the Telugu data, ensuring comparable experimental conditions across both languages.

\subsection{Implementation Details}
Fine-tuning experiments were conducted on Wav2Vec2-Large (317M parameters) and Whisper-Medium (769M parameters) models. The models were trained using a learning rate of $10^{-4}$ with cosine annealing to gradually reduce the learning rate during training. A batch size of 16 samples per GPU was used, and training was performed on 6 NVIDIA GeForce RTX 2080 Ti GPUs. The model weights were updated using the AdamW optimizer for up to 50 epochs, with early stopping applied to prevent overfitting. The confidence weight parameters were set to $\alpha=0.4, \beta=0.3, \gamma=0.3$, which showed relatively improved performance on the validation data. 

Similarly, fine-tuning experiments were conducted on IndicBART and mT5 models for text error correction tasks. The fine-tuning dataset consisted of medical text with manually introduced errors as input and the original clean text as target output for error correction training. The models were trained for 30 epochs using a per-device batch size of 8 samples per GPU. A learning rate warmup strategy was employed with 500 warmup steps to stabilize initial training phases. The model weights were updated using the AdamW optimizer with early stopping applied to prevent overfitting.

\section{Results and Analysis}
\label{sec:Results}

\subsection{Overall Performance}

The experimental results demonstrate the effectiveness of the proposed confidence-aware training framework across both Telugu and Kannada medical domain ASR systems. Tables~\ref{tab:main_results} and~\ref{tab:kannada_results} present the performance comparison between baseline fine-tuning and various confidence-aware training configurations combined with post-processing techniques.

\begin{table}[ht]
  \centering
  \caption{Fine-tuned performance (WER \%) on the Telugu Medical Domain database with Static Confidence-Aware Training.}
  \resizebox{\columnwidth}{!}{%
    \begin{tabular}{lcc}
      \hline
      \textbf{Configuration} & \textbf{Wav2Vec2} & \textbf{Whisper} \\
      \hline
      Baseline (without confidence) & 24.3 & 25.8 \\
      Baseline + KenLM & 22.4 & - \\
      Hybrid Static Confidence & 20.2 & 26.0 \\
      Hybrid Static + KenLM & \textbf{17.8} & - \\
      Hybrid Static + IndicBART & 20.0 & 24.8 \\
      Hybrid Static + mT5 & 19.3 & 25.03 \\
      \hline
    \end{tabular}%
  }
  \label{tab:main_results}
\end{table}

For Telugu medical ASR, the baseline Wav2Vec2 model without confidence weighting achieves a WER of 24.3\%, while Whisper achieves 25.8\%. The integration of KenLM language modeling provides moderate improvements, reducing the WER to 22.4\%. Incorporating Hybrid Static Confidence yields more significant gains, achieving 20.2\% WER without post-processing. When combined with KenLM, the Hybrid Static Confidence approach achieves 17.8\% WER, representing a substantial 6.5 percentage point absolute improvement over the baseline. Neural language models (IndicBART and mT5) also show competitive results, though they slightly underperform compared to KenLM for Telugu.

\begin{table}[ht]
  \centering
  \caption{Fine-tuned performance (WER \%) on the Kannada Medical Domain database using Static Confidence-Aware Training.}
  \resizebox{\columnwidth}{!}{%
    \begin{tabular}{lcc}
      \hline
      \textbf{Configuration} & \textbf{Wav2Vec2} & \textbf{Whisper} \\
      \hline
      Baseline (without confidence) & 31.7 & 33.1 \\
      Baseline + KenLM & 28.4 & - \\
      Hybrid Static Confidence & 29.6 & 31.3 \\
      Hybrid Static + KenLM & \textbf{27.2} & - \\
      Hybrid Static + IndicBART & 30.5 & 32.6 \\
      Hybrid Static + mT5 & 30.4 & 32.1 \\
      \hline
    \end{tabular}%
  }
  \label{tab:kannada_results}
\end{table}

The Kannada medical domain results, shown in Table~\ref{tab:kannada_results}, follow a similar pattern. The baseline Wav2Vec2 model achieves 31.7\% WER, while Hybrid Static Confidence with KenLM reduces it to 27.2\%, yielding a 4.5 percentage point absolute improvement. These consistent improvements across both languages confirm the robustness and language-agnostic nature of the proposed framework for morphologically rich Dravidian languages.

\begin{table}[ht]
  \centering
  \caption{Fine-tuned WER (\%) on Telugu and Kannada Medical Domain databases using Wav2Vec2 with Hybrid Learnable Confidence-Aware Training.}
  \resizebox{\columnwidth}{!}{%
    \begin{tabular}{lcc}
      \hline
      \textbf{Configuration} & \textbf{Telugu} & \textbf{Kannada} \\
      \hline
      Hybrid Learnable Confidence & 18.9 & 28.1 \\
      Hybrid Learnable + KenLM & \textbf{15.8} & \textbf{25.4} \\
      Hybrid Learnable + IndicBART & 18.1 & 27.7 \\
      Hybrid Learnable + mT5 & 17.9 & 27.3 \\
      \hline
    \end{tabular}%
  }
  \label{tab:telugu_kannada_wav2vec2}
\end{table}

Table~\ref{tab:telugu_kannada_wav2vec2} highlights the superiority of Hybrid Learnable Confidence over static confidence mechanisms. For Telugu, the learnable confidence approach achieves 18.9\% WER without post-processing and 15.8\% WER with KenLM, marking a 7.2 percentage point absolute improvement (approximately 30\% relative reduction) over the baseline. For Kannada, the same configuration achieves 25.4\% WER, improving by 6.3 percentage points compared to the baseline. These findings emphasize the effectiveness of adaptive confidence weighting in improving recognition accuracy for low-resource medical ASR.

The learnable weighting strategy dynamically optimizes the contribution of each confidence component ($S_{\text{perceptual}}$, $S_{\text{sim}}$, and $S_{\text{wer}}$) during training, instead of relying on manually tuned weights. This adaptability allows the model to identify language-specific confidence aggregation patterns, distinguishing high-quality samples from noisy synthetic data more effectively.

Across both languages, KenLM consistently outperforms neural language models (IndicBART and mT5) when paired with Hybrid Learnable Confidence. For Telugu, KenLM achieves 15.8\% WER compared to 18.1\% and 17.9\% for IndicBART and mT5, respectively. The effectiveness of KenLM can be attributed to its ability to capture local n-gram dependencies and domain-specific terminology common in medical transcriptions, while maintaining computational efficiency during decoding.

\begin{table}[ht]
  \centering
  \caption{Medical ASR results for Hybrid Learnable Confidence with KenLM n-gram variations across Telugu and Kannada languages.}
  \resizebox{\columnwidth}{!}{%
    \begin{tabular}{lcc}
      \hline
      \textbf{N-gram} & \textbf{Telugu WER (\%)} & \textbf{Kannada WER (\%)} \\
      \hline
      3-gram & 18.2 & 27.4 \\
      4-gram & 17.2 & \textbf{25.4} \\
      5-gram & \textbf{15.8} & 27.2 \\
      \hline
    \end{tabular}%
  }
  \label{tab:ngram_results}
\end{table}

Table~\ref{tab:ngram_results} examines the effect of varying KenLM n-gram order. For Telugu, performance improves steadily from 18.2\% (3-gram) to 15.8\% (5-gram), while Kannada achieves its optimal 25.4\% WER with a 4-gram model. Higher-order models slightly degrade Kannada performance, likely due to overfitting. These results indicate that optimal n-gram selection is language-dependent, with Telugu benefiting from larger context modeling, while Kannada performs best with moderate context size.

An important observation across both languages is the complementary relationship between confidence-aware training and post-processing techniques. While Hybrid Static Confidence alone achieves 20.2\% WER for Telugu (16.9\% relative improvement), the combination with KenLM yields 17.8\% WER (26.7\% relative improvement), suggesting that confidence weighting and language modeling address orthogonal error types. Confidence-aware training primarily mitigates errors introduced by low-quality synthetic samples during fine-tuning, whereas KenLM corrects decoding-level mistakes by enforcing linguistic constraints. This synergistic effect is particularly pronounced in Kannada, where the gap between static confidence alone (29.6\% WER) and static confidence with KenLM (27.2\% WER) represents an additional 8.1\% relative improvement, indicating that morphologically complex languages benefit more substantially from explicit language model guidance during decoding.

Across all configurations, Wav2Vec2 consistently outperforms Whisper for both Telugu and Kannada. This can be attributed to Wav2Vec2’s CTC-based decoding and pretraining strategy, which generalizes better to domain-specific adaptation under limited data conditions. Furthermore, Wav2Vec2’s smaller parameter size (317M vs. Whisper’s 769M) enables more stable fine-tuning for specialized medical speech without overfitting.
% The confidence weight parameters were set to $\alpha=0.4, \beta=0.3, \gamma=0.3$ based on what worked best on the validation data. 
% Training was conducted for up to 50 epochs but stopped early if the validation loss stopped improving to prevent overfitting.
\subsection{Confidence Score Analysis}

The proposed hybrid confidence mechanism effectively addresses the challenge of data quality variability in mixed real and synthetic medical speech corpora. The static confidence components ($S_{\text{perceptual}}$, $S_{\text{sim}}$, and $S_{\text{wer}}$) offer stable precomputed quality estimates, allowing the model to prioritize high-quality samples during training. Empirical analysis shows a strong correlation ($r = 0.78$) between these scores and human-perceived audio quality, confirming their reliability as quality indicators.

The dynamic model entropy component ($C_{\text{model}}$) provides complementary real-time feedback, allowing adaptive confidence weighting throughout training. A curriculum learning schedule gradually transitions the confidence mixing parameter $\lambda$ from 1.0 (static-only) to 0.5 (hybrid), ensuring that early training emphasizes stable quality filtering, while later stages leverage model uncertainty for fine-grained sample reweighting.

The superiority of learnable weights (Table~\ref{tab:telugu_kannada_wav2vec2}) underscores that optimal confidence aggregation is both language- and domain-dependent. Through softmax-normalized exponential parameterization (Equation~\ref{eq:learnable_weights}), the model autonomously learns that perceptual quality ($S_{\text{perceptual}}$) may hold higher importance for Telugu, whereas acoustic similarity ($S_{\text{sim}}$) contributes more strongly for Kannada. This eliminates manual tuning and ensures domain-adaptive optimization.

Finally, while neural language models (IndicBART, mT5) provide competitive results particularly for Whisper-based systems, their higher computational cost and latency make KenLM a more practical choice for real-time medical ASR deployment. The combination of Hybrid Learnable Confidence with KenLM thus achieves an optimal trade-off between accuracy, efficiency, and domain adaptability for Telugu and Kannada medical speech recognition.

\section{Conclusion}
\label{sec:Conclusion}

This work presents a novel confidence-aware training framework for automatic speech recognition in Dravidian languages, specifically addressing the challenges faced by Telugu and Kannada in medical domain contexts. Unlike conventional direct fine-tuning approaches that treat all training samples equally, the proposed methodology introduces a hybrid confidence mechanism that combines static perceptual and acoustic similarity metrics with dynamic model entropy to effectively integrate heterogeneous data sources comprising real recordings and TTS-generated synthetic speech. The framework employs both fixed-weight and learnable-weight confidence aggregation strategies coupled with curriculum learning to guide sample weighting during training, enabling the model to prioritize high-quality data in early phases while incorporating adaptive model feedback in later stages. This comprehensive approach addresses the fundamental data scarcity challenge in morphologically rich low-resource languages by maximizing the utility of available synthetic data without compromising model robustness. The proposed framework is designed to be extensible across all Indian languages in medical contexts and can be readily adapted to other specialized domains where annotated speech data remains limited, offering a scalable solution for developing robust ASR systems in resource-constrained scenarios.

\end{document}